# Site-Selective Activation of Near-Infrared Emitters in hBN

Tadas Paulauskas,[1,*] Vakaris Šilys,[1] Edgaras Markauskas,[1] Julius Janušonis,[1] Ifra Bibi,[1] Vitalij Kovalevskij,[1] Skirmantas Keršulis,[1] Danielis Rutkauskas[1] and Marek Maciaszek[1,2]

[1] State Research Institute Center for Physical Sciences and Technology, Saulėtekio al. 3, Vilnius LT-10257, Lithuania

[2] Faculty of Physics, Warsaw University of Technology, Koszykowa 75, 00-662 Warsaw, Poland

[*] E-mail: tadas.paulauskas@ftmc.lt

## Abstract

Near-infrared (NIR) emitters in hexagonal boron nitride (hBN) have recently emerged as bright and spectrally stable light sources, but they have so far been found at locations set by the local defect environment rather than by design. Here we show that NIR emitters can be activated at predefined sites by combining single-shot femtosecond laser writing with annealing in an oxygen-rich environment. To our knowledge, this is the first demonstration of site-selective activation of NIR emitters in hBN. The written matrices are recovered in post-anneal photoluminescence maps, including in collection bands centred at 800 and 900 nm. Site-resolved spectra under 532, 635 and 785 nm excitation show narrow lines extending to 1 μm; 83% of the spectra acquired at written sites under 785 nm excitation and 71% under 635 nm excitation contain at least one resolved peak above 810 nm. The integrated emission increases with writing-pulse energy over the range studied, and many sites host more than one emitter, which identifies lower writing energies as the route to single-emitter occupancy. Photon-correlation measurements on selected sites show antibunching, with a raw, uncorrected second-order correlation value $g^{(2)}(0)$ of 0.31 in a band centred at 850 nm. Spectral time series reveal bands that persist over the recorded intervals together with intermittent emission beyond 900 nm. Comparison with identically written, argon-annealed samples indicates that the annealing atmosphere controls the spectral range of the activated emitters, while the laser defines where they form.

## 1. Introduction

Single-photon emitters in hexagonal boron nitride (hBN) combine room-temperature operation with the mechanical flexibility and heterogeneous-integration capabilities of a van der Waals material. Optically active defects in hBN have been observed across an unusually broad spectral range, from the ultraviolet to the near-infrared (NIR) [1]. At the same time, microscopic identification of many of these optical signatures remains challenging [2]. This spectral diversity creates opportunities for interfacing hBN emitters with different photonic platforms, but practical integration also requires control over where the emitters form. An emitter generated at an arbitrary position is unlikely to coincide with the optical mode of a resonator, waveguide or collection structure.

Spatial control has therefore become a central fabrication problem. Femtosecond irradiation can directly generate optically active defects and spin ensembles, with outcomes that depend strongly on dose and pulse energy [3]. Laser processing is particularly attractive because it is maskless, programmable and capable of modifying hBN on submicrometre length scales. Single-pulse irradiation followed by thermal treatment has produced visible single-photon-emitter arrays with a reported single-emitter yield of 43% under optimized conditions [4], while related approaches have generated visible spin-defect arrays [5], anneal-activated laser-written emitters [6] and emitters integrated with SiN photonic structures [7]. Near the breakdown threshold, single-shot irradiation can additionally produce subwavelength modification of hBN [8]. Electron-beam activation [9], focused-ion-beam processing [10] and nanoindentation-assisted functionalization [11] further show that localized lattice modification provides a spatial handle for defect formation. These demonstrations of controlled placement, however, concern visible emitters; site-selective fabrication of NIR-emitting centres has remained undeveloped [12–15].

Oxygen-rich processing addresses the spectral side of the problem. The density and optical properties of emitters in exfoliated hBN depend strongly on oxygen flow and annealing temperature [16], and oxygen-based treatments have produced narrow emission in the 700–820 nm range [17], sharp zero-phonon lines (ZPLs) with reduced broadband fluorescence [18] and NIR emitters with optically addressable spin signatures [19]. Oxygen-plasma treatment followed by annealing has further yielded bright quantum emitters extending to 971 nm [20]. These results establish oxygen availability as a key processing parameter for NIR-active defects. In all of these studies, however, the emitters were found at locations determined by the local defect environment rather than at predefined coordinates. The NIR emitters that currently attract the most interest are therefore the ones for which no placement method exists.

Here we address this gap by combining the two controls. Building on our previous single-pulse laser-writing study, in which argon annealing produced emitters predominantly in the visible range [21], we write matrices of single-pulse modification sites and anneal the samples at high temperature in pure oxygen. The written geometry is recovered in post-anneal photoluminescence (PL) maps, including in narrow collection bands centred at 800 and 900 nm, and site-resolved spectra show narrow emission lines extending to approximately 1 μm at the large majority of written sites. To our knowledge, this is the first demonstration that NIR emitters in hBN can be activated at predefined positions. We characterize the resulting emitter population through multiwavelength spectroscopy at matched sites, peak statistics, pulse-energy dependence, time-resolved PL, photon correlations and spectral time series. The present process yields more than one emitter at many sites, and single-photon emission is verified at selected sites rather than across the array; we therefore also identify the parameters that should bring the process to single-emitter occupancy.

## 2. Methods

### 2.1. Sample preparation

Bulk hBN crystals (HQ Graphene) were mechanically exfoliated with adhesive tape onto 0.5 mm fused-silica substrates (PI-KEM), following the route used in our argon-annealed study [21]. Before exfoliation, the substrates were cleaned sequentially in acetone, isopropanol and deionized water and treated with oxygen plasma. The exfoliated samples were baked in air at 350 °C for 30 min to reduce tape residue.

### 2.2. Femtosecond laser writing

Laser writing used a 1030 nm femtosecond source (Pharos, Light Conversion; 250 fs pulse duration) with an internal pulse picker delivering one pulse per site [21]. The near-Gaussian, linearly polarized beam was attenuated to the selected pulse energy and focused under ambient conditions through a 50× objective (NA 0.65) to a spot of approximately 1.6 μm. The axial focus was referenced to a nearby flake edge. Sites were written as 10 × 10 matrices with a 5 μm pitch, with the pulse energy varied by column over 44.7, 46.5, 48.0, 49.6, 51.2, 52.7, 54.3, 55.9, 57.0 and 58.2 nJ. The two matrices analysed here were written on the same large flake, approximately 300 nm thick. An optical image of a written matrix is shown in figure S3.

### 2.3. Oxygen annealing

After laser writing, the samples were annealed in pure $O_2$ in the same tube furnace (Carbolite GHA 12/600) used for the earlier argon-annealed samples [21]. The $O_2$ flow was approximately 200 mL

min$^{-1}$, with a 20 °C min$^{-1}$ ramp to 1000 °C, a 60 min dwell and cooling to 200 °C before the samples were removed.

### 2.4. Confocal PL mapping and spectroscopy

All optical measurements were performed at room temperature. Spectra and the overview map were acquired on a custom inverted confocal microscope (Nikon Eclipse Ti-U, 40×/0.60 objective, piezoelectric scanning stage, 75 μm pinhole) described previously [21]. Emission was detected with a single-photon avalanche diode for mapping (0.18 μm pixels, 1 ms dwell time) or dispersed in a spectrometer (Shamrock SR-303i, 50 lines mm$^{-1}$ grating) onto an EMCCD camera. Excitation at 532, 635 and 785 nm used continuous-wave diode-pumped solid-state lasers with nominal powers at the sample of 1, 0.3 and 0.6 mW, and long-pass filters at 550, 650 and 800 nm, respectively. Spectra were acquired with integration times of 1 s (532 nm), 0.5 s (635 nm) and 1 or 10 s (785 nm); the 635 and 785 nm spectra were accumulated over 10 frames. The band-pass-filtered maps in figures 1(b)–(g) were acquired on a second confocal microscope (60×/0.90 objective, 50 μm pinhole) with a 640 nm diode laser (PicoQuant LDH-D-C-640) [21].

### 2.5. Spectral statistics

One spectrum was analysed per measured written site in each excitation series, giving 159 spectra under 635 nm excitation and 132 under 785 nm excitation. Peaks were identified in baseline-corrected, lightly smoothed spectra within the 812–1020 nm window using fixed prominence, height, width and separation thresholds referenced to the noise level; a spectrum was classified as NIR-positive when it contained at least one accepted peak. Under 785 nm excitation, the hBN Raman line near 879 nm was treated separately by local fitting so that it was not counted as emission. The full criteria are given in section S1 of the supplementary material. Pulse-energy-resolved intensities are reported as the mean of the normalized integrated spectral intensity with the standard error of the mean, and fractions are quoted with 95% Wilson confidence intervals.

### 2.6. Time-resolved PL and photon correlations

Time-resolved PL and second-order correlations were measured on the second confocal microscope using the 640 nm diode laser in pulsed (10 MHz) and continuous-wave mode, respectively. Emission was selected with 40 nm band-pass filters centred at 800, 850 and 900 nm. Decays were recorded by time-correlated single-photon counting (PicoHarp 300) and fitted with a biexponential model. For correlations, the emission was divided by a 50:50 beam splitter onto two single-photon counting modules (Excelitas SPCM-AQRH-14); histograms had a native bin width of 0.512 ns and were normalized to the long-delay coincidence level. The traces were fitted with an exponential antibunching term plus one bunching term, with the model averaged over the experimental bins. No background correction was applied. Fit models and parameters are given in section S4 of the supplementary material.

### 2.7. Spectral time series

Spectral time series were recorded under 785 nm excitation with a 0.3 s frame time. For each site in figure 6, a 60 s sequence and a 180 s sequence acquired approximately one hour later were wavelength-calibrated, interpolated to a common axis, intensity-aligned and displayed consecutively, giving 240 s of cumulative acquisition.

## 3. Results

### 3.1. Preferential activation at laser-written sites

Following $O_2$ annealing, the laser-written matrices appear as regularly spaced PL maxima (figure 1). Both the overview map acquired under 532 nm excitation and the approximately 16 × 16 µm maps acquired under 640 nm excitation reproduce the imposed 5 µm pitch and matrix geometry. This correspondence between the designed coordinates and the post-anneal PL maxima identifies the laser-modified regions as preferential activation sites.

The maps in figures 1(b)–(g) sample representative regions through different collection windows: a 675–900 nm band-pass filter (b)–(d), a 40 nm band-pass filter centred at 800 nm (e) and (f) and a 40 nm band-pass filter centred at 900 nm (g). The matrix remains clearly visible in each window, including the narrow 900 nm band, which shows directly that the spatially selective activation extends into the NIR. The maps are displayed on a logarithmic scale so that weaker features are visible. Some PL spots do occur away from the written coordinates, most noticeably near pre-existing structural damage such as the fractured region in figure 1(d); these are consistent with activation at pre-existing or processing-induced defects during annealing. Because the panels represent different regions and collection windows, they are not used for a quantitative comparison of off-site emitter density between bands.

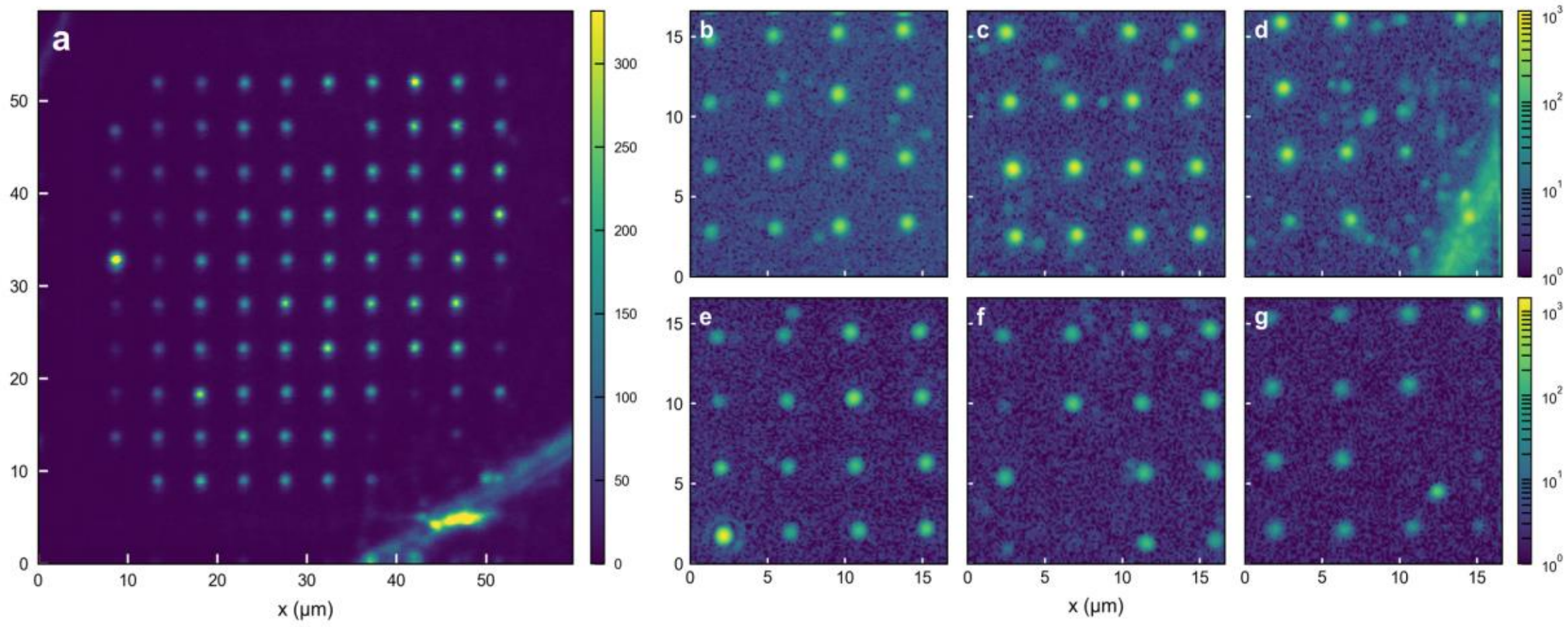

**Figure 1.** Preferential PL activation at laser-written coordinates after oxygen annealing. (a) Overview confocal PL map of a written matrix under 532 nm excitation, collected through a 550 nm long-pass filter and shown on a linear intensity scale. The writing pulse energy increases column-wise from left to right: 44.7, 46.5, 48.0, 49.6, 51.2, 52.7, 54.3, 55.9, 57.0 and 58.2 nJ. (b)–(g) Representative confocal maps from written regions under 640 nm excitation, shown on a logarithmic intensity scale and collected through a 675–900 nm band-pass filter (b)–(d), a 40 nm band-pass filter centred at 800 nm (e) and (f) and a 40 nm band-pass filter centred at 900 nm (g).

### 3.2. NIR spectral landscape

The written sites show pronounced emission across a broad spectral range, with resolved features extending deep into the NIR (figure 2). Under 635 nm excitation, representative spectra contain prominent narrow peaks up to approximately 940 nm (figures 2(a)–(c)). Individual sites range from spectra dominated by a single narrow line to multipeak emission. Recurring features are observed near 700 nm, around 800–820 nm and close to 900 nm. In particular, narrow ZPL-like peaks near 900 nm occur at multiple written sites and remain efficiently accessible under red excitation.

Excitation at 785 nm selectively probes the long-wavelength population and reveals resolved emission from approximately 800 nm to the detection limit near 1 μm (figure 2(d)). Compared with 635 nm excitation, emission in the 820–870 nm range becomes considerably more prominent, together with recurring peaks near 900 nm. Because the sensitivity of the silicon detector decreases strongly towards 1 μm, the measured spectra probably under-represent the longest-wavelength emission.

The spectral diversity indicates that the process accesses a family of NIR-active configurations rather than one spectrally uniform centre; differences in microscopic configuration, charge state, local strain or defect environment may all contribute. The occurrence of both single-peak-dominated and multipeak spectra also shows that the number of active centres varies between written sites, so that a bright written spot cannot by itself be taken as a single emitter. Spatial mapping is therefore combined below with spectral statistics and photon-correlation measurements.

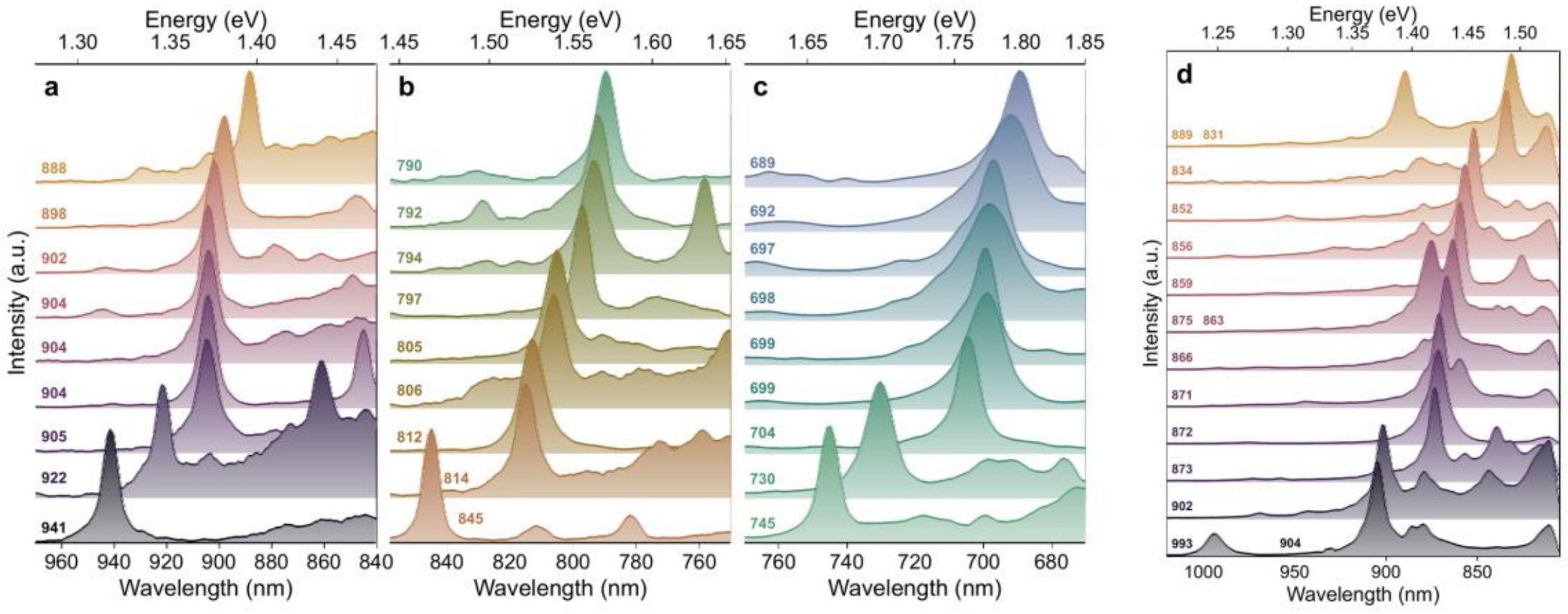

**Figure 2.** Representative room-temperature spectra from laser-written sites. (a)–(c) Vertically offset, normalized spectra from distinct written sites under 635 nm excitation, grouped into three spectral regions. The sites span several writing energies; numeric labels indicate prominent peak positions in nanometres. (d) Long-wavelength spectra from distinct sites under 785 nm excitation.

### 3.3. Multiwavelength excitation at matched sites

Spectra acquired at the same written coordinates under two or three excitation wavelengths (532, 635 and 785 nm) show how the response depends on excitation wavelength (figure 3). Excitation at 532 nm accesses a broad visible-to-NIR range and reveals additional short-wavelength features, including pronounced emission near 575 nm at many written sites. Excitation at 635 nm emphasizes emission above the 650 nm long-pass cut-off, whereas 785 nm excitation selectively addresses the population above approximately 800 nm.

Several spectral features recur at the same coordinates under different excitation conditions. Prominent emission near 700 nm is observed under both 532 and 635 nm excitation, while narrow features near 800 nm recur at several sites (figures 3(a), S2(d) and S2(e)) and are generally more pronounced under 635 nm excitation. Small variations in the positions of otherwise similar ZPL-like features are consistent with the sensitivity of hBN defect transitions to their local environment, including lattice strain [22].

The strongest excitation dependence is observed for the long-wavelength features. At several sites, peaks that are clearly resolved under 785 nm excitation are weak or absent under 635 nm excitation (for example, figures 3(a) and (c)). This is consistent with excitation-dependent access to co-located centres, arising for instance from differences in absorption cross-section or from photoinduced charge-state dynamics. The excitation wavelength thus provides an additional degree of selectivity within the emission generated at a single written site. Because excitation powers and system responses differ between measurements, the normalized spectra are used to compare peak positions and line shapes, not absolute intensities. Additional matched-site spectra are shown in figure S2.

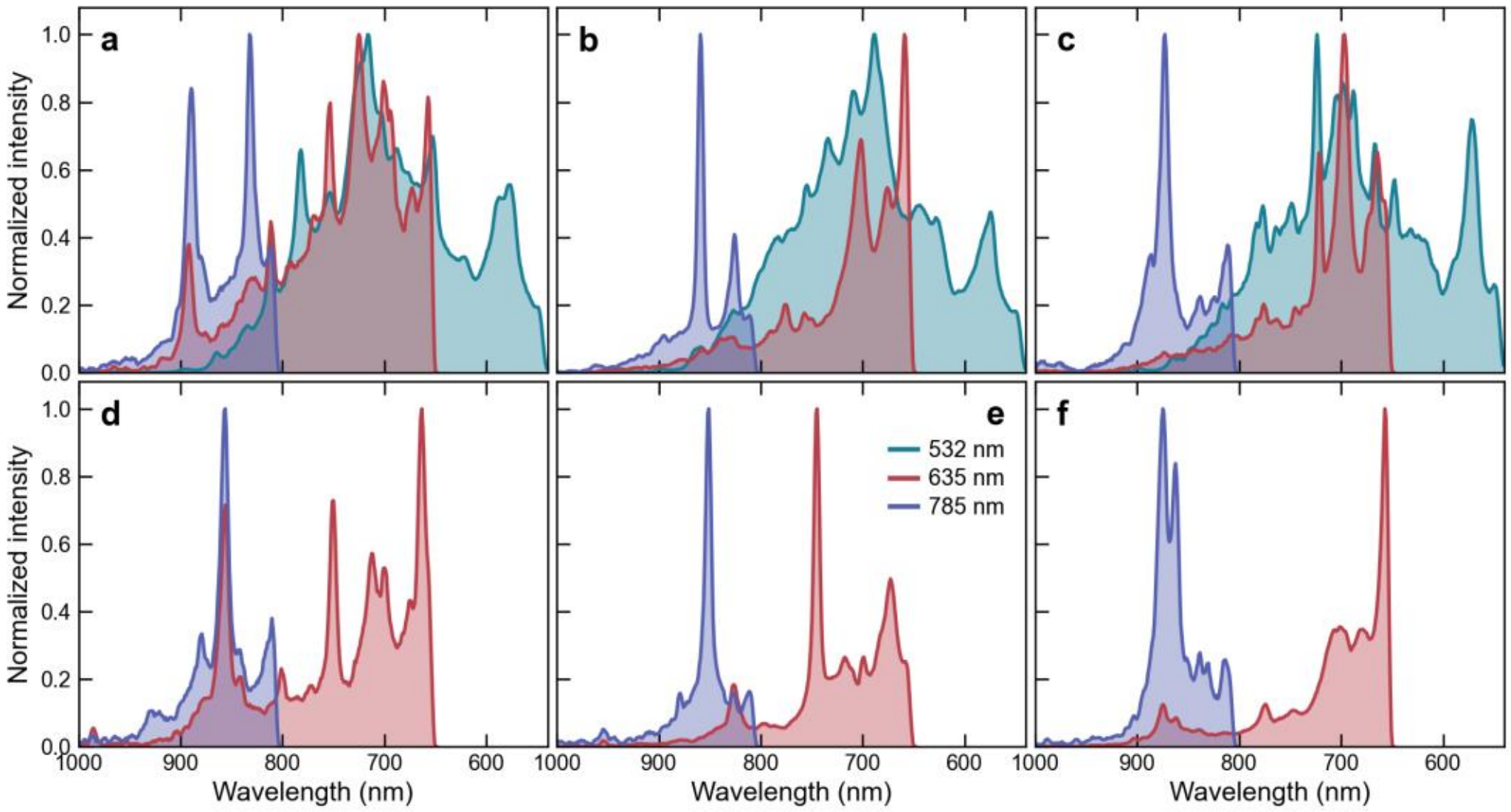

**Figure 3.** Excitation-wavelength dependence at matched laser-written sites. (a)–(c) Normalized spectra from three coordinates in the first written matrix measured under 532 nm (green), 635 nm (red) and 785 nm (blue) excitation. (d)–(f) Normalized spectra from three coordinates in the second written matrix measured under 635 and 785 nm excitation. Within each panel, the spectra correspond to the same coordinate. The sites span several writing energies.

### 3.4. NIR prevalence, multiplicity and pulse-energy dependence

Figure 4(a) shows spectra averaged over the written sites. Under 532 nm excitation, the averaged spectrum has maxima near 575 and 700 nm and decreases rapidly above 800 nm. Under 635 nm excitation, strong emission remains centred near 700 nm, with the shorter-wavelength response truncated by the long-pass filter. At longer wavelengths, the averaged 635 and 785 nm spectra decrease gradually but retain weaker features associated with recurring bands in the site-resolved spectra, for example near 870 nm.

NIR emission is the rule rather than the exception at the written sites. Taking 810 nm as an operational lower boundary, 110 of 132 spectra acquired under 785 nm excitation contain at least one resolved NIR peak (83%; 95% confidence interval 76–89%), as do 113 of 159 spectra under 635 nm excitation (71%; 64–78%). These numbers are the fractions of measured written sites showing resolved NIR emission; they are not single-emitter yields. The resolved peaks are distributed mainly between 810 and 910 nm, with a tail extending to 1 μm (figure 4(c)).

The peak multiplicity is summarized in figure 4(d). Under 785 nm excitation, 22 spectra contain no resolved NIR peak, 54 contain one, 38 contain two and 18 contain three or more; under 635 nm excitation the corresponding counts are 46, 74, 30 and 9. Among NIR-positive spectra, 49% (785 nm) and 65% (635 nm) therefore contain a single resolved NIR peak. The larger multiplicity seen under 785 nm excitation is consistent with selective access to long-wavelength features that are weak or unresolved under red excitation. A substantial fraction of the written sites is thus already single-peak-dominated in the NIR under the present conditions, while the remainder host several emitters.

The normalized integrated spectral intensity increases overall with writing-pulse energy across the investigated range (figure 4(b)), by roughly 50% under 635 nm excitation and 20% under 785 nm excitation between the lowest and highest energies. At the highest pulse energies the trend is perturbed by an extended structural defect that intersects part of the matrix and reduces or quenches the emission from several sites, which is most apparent in the 532 nm data. Variations in local thickness, pre-existing disorder and laser focus contribute additional site-to-site scatter. The pulse-energy window explored here is narrow and lies entirely in the regime where most sites are activated, so the trend should be read as an indication of the direction of control: lower writing energies, closer to the modification threshold, are expected to reduce the number of emitters per site.

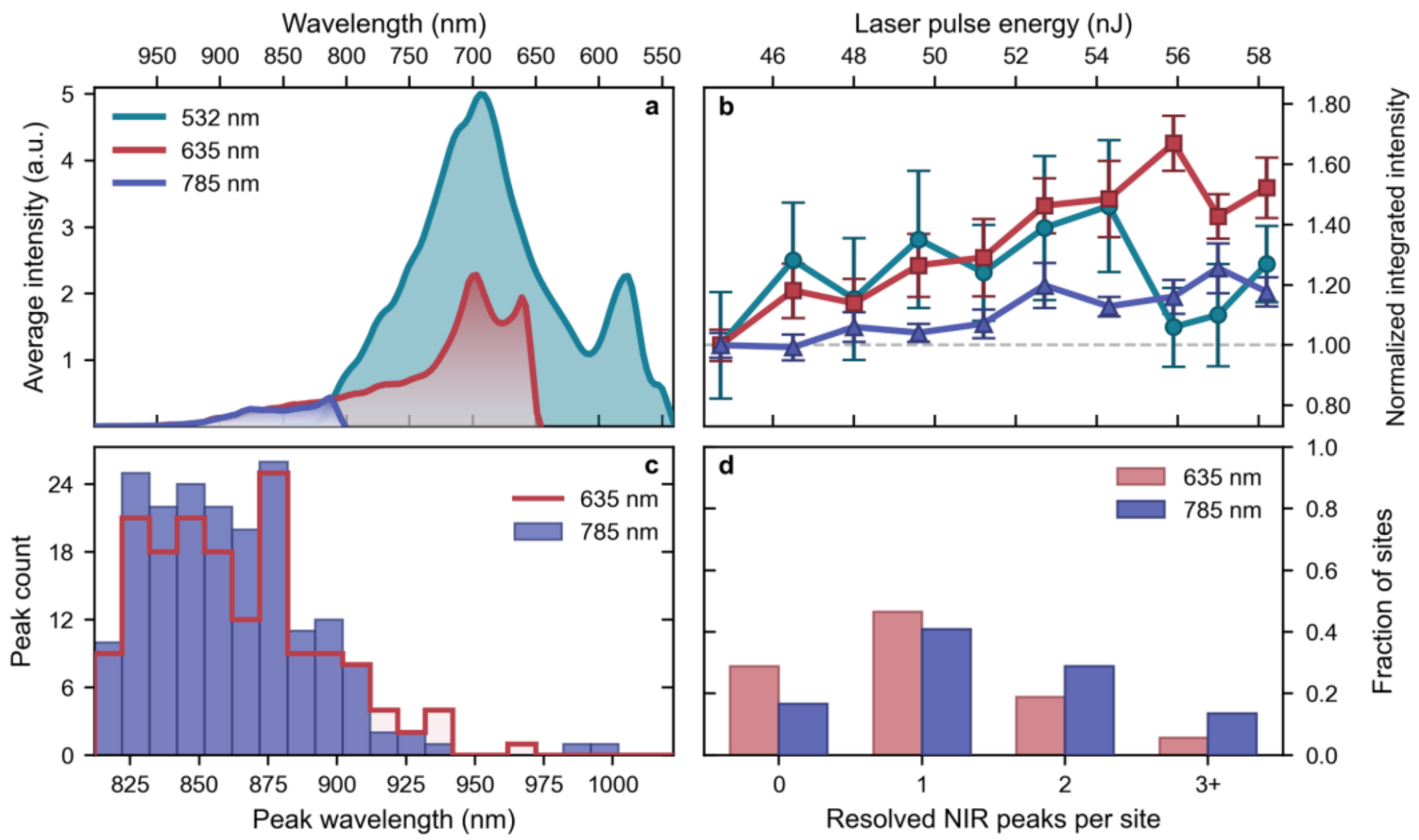

**Figure 4.** Ensemble spectral statistics and pulse-energy dependence. (a) Average spectra under 532, 635 and 785 nm excitation. (b) Normalized integrated intensity versus single-pulse writing energy. Points show the mean and error bars the standard error of the mean; colours as in (a). (c) Distribution of resolved peak wavelengths above 810 nm under 635 and 785 nm excitation. (d) Fraction of analysed spectra containing zero, one, two, or three or more resolved NIR peaks.

### 3.5. Excited-state dynamics and photon statistics

Time-resolved PL and second-order photon-correlation measurements were performed under 640 nm excitation on selected written sites showing single-peak-dominated emission, with collection in 40 nm bands centred at 800, 850 and 900 nm (figure 5). The decays yield intensity-weighted mean lifetimes of approximately 1.9 ns in the 800 nm band, 3.7 ns in the 850 nm band and 1.7 ns in the 900 nm band. The 850 nm decay is also reasonably described by a single exponential, whereas the 800 and 900 nm decays require two components. Component-resolved fit parameters are given in table S1.

All three correlation traces show a pronounced zero-delay antibunching dip together with bunching at longer delays. The raw fitted minima are $g^{(2)}(0)$ = 0.50, 0.31 and 0.54 for the 800, 850 and 900 nm bands, respectively. The 850 nm value lies well below 0.5 and demonstrates single-photon emission from a laser-defined NIR-active site. The 800 and 900 nm values lie at the 0.5 boundary. All three values are conservative, for two reasons. First, no background correction was applied, and each 40 nm collection window admits spectral background and the tails of neighbouring emission features. Second, the 0.512 ns bin width is comparable to the antibunching time constants (1.3, 3.1 and 1.6 ns for the 800, 850 and 900 nm bands), so the reported minima are averages over the central bin. Evaluated at zero delay, the fitted model gives approximately 0.25 for the 850 nm band and 0.38 for the 800 nm band; for the 900 nm band the bunching term is too poorly constrained at short delays for a meaningful extrapolation (table S2).

The antibunching time constants are comparable to the corresponding PL decay times. The longer-delay bunching in all three traces indicates additional population dynamics, consistent with shelving in one or more metastable states beyond a two-level description.

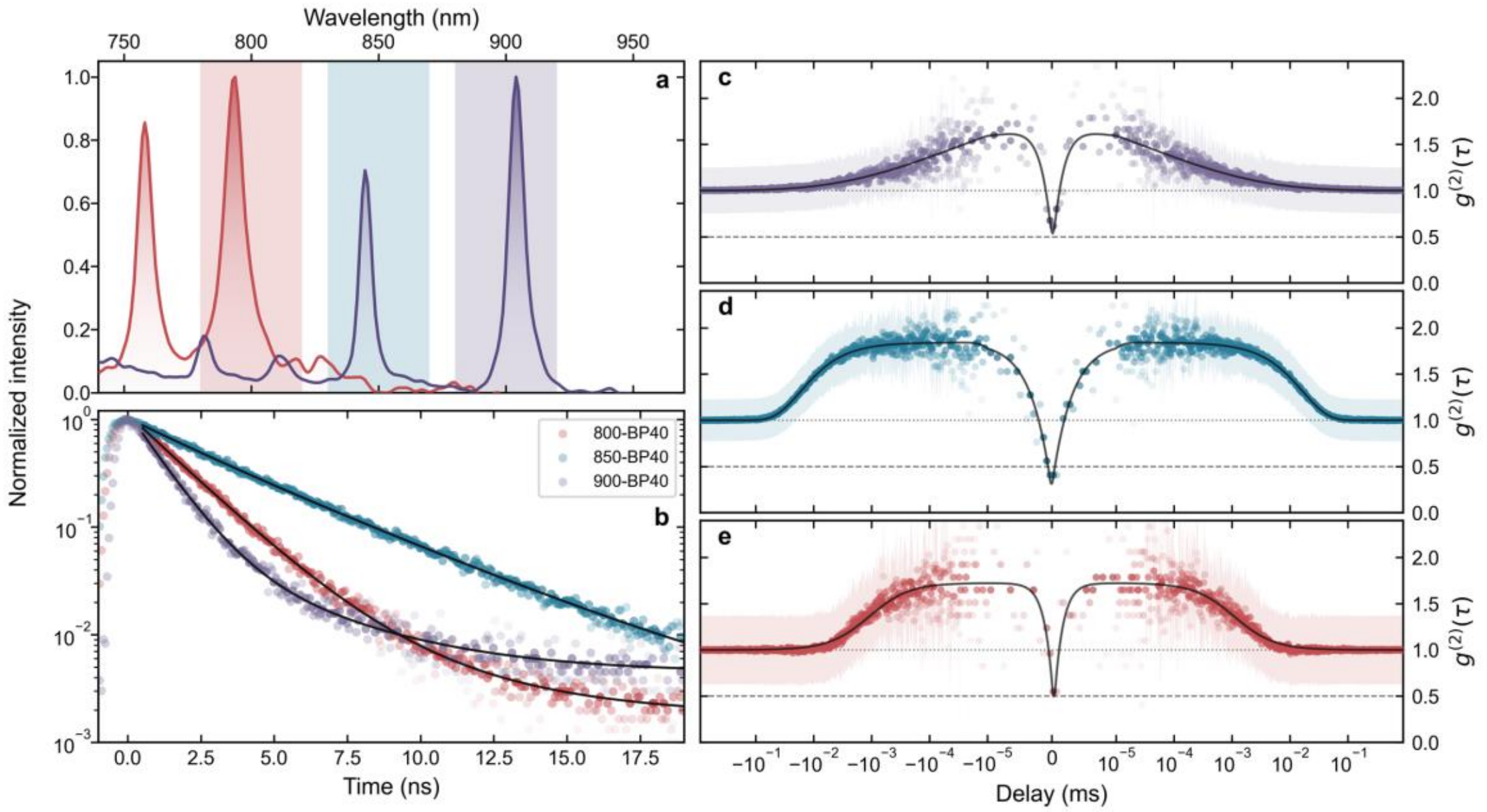

**Figure 5.** Time-resolved PL dynamics and photon correlations. (a) Representative spectra with the 40 nm collection windows centred at 800, 850 and 900 nm shaded. (b) Normalized time-resolved PL traces with monoexponential (850 nm band) and biexponential (800 and 900 nm bands) fits. (c)–(e) Continuous-wave second-order correlation functions with fitted antibunching and bunching models for the 900, 850 and 800 nm bands, respectively; shaded bands show one standard deviation. The measurements were selected for band-resolved dynamics and do not form a pulse-energy series.

### 3.6. Spectral stability

Spectrally resolved time series acquired under 785 nm excitation reveal both persistent and intermittent NIR emission from the written sites (figure 6). Each panel combines two sequences recorded approximately one hour apart at the same site. Several bands in the 800–900 nm range remain at fixed spectral positions throughout, and additional measurements show persistent emission near 950 nm (figure S1). This behaviour is in line with the spectral stability reported for long-wavelength hBN emitters created by oxygen processing [20], although microscopic identification would be needed to establish whether the same defect configurations are involved.

Alongside these persistent bands, several long-wavelength features are intermittent. In figure 6(a), a feature near 945 nm switches on only for brief intervals and a feature near 850 nm appears twice for about a second. A band near 950 nm in figure 6(b) shows a different switching pattern with longer on-periods, a feature near 980 nm in figure 6(c) switches on once during the recording, and a feature near 920 nm in figure 6(d) is intermittent throughout. Intermittency of this kind may reflect charge-state switching, population transfer involving metastable states or interaction with the local defect environment. Its occurrence at wavelengths approaching 1 µm is notable, and a quantitative account will require measurements on isolated emitters under controlled excitation conditions.

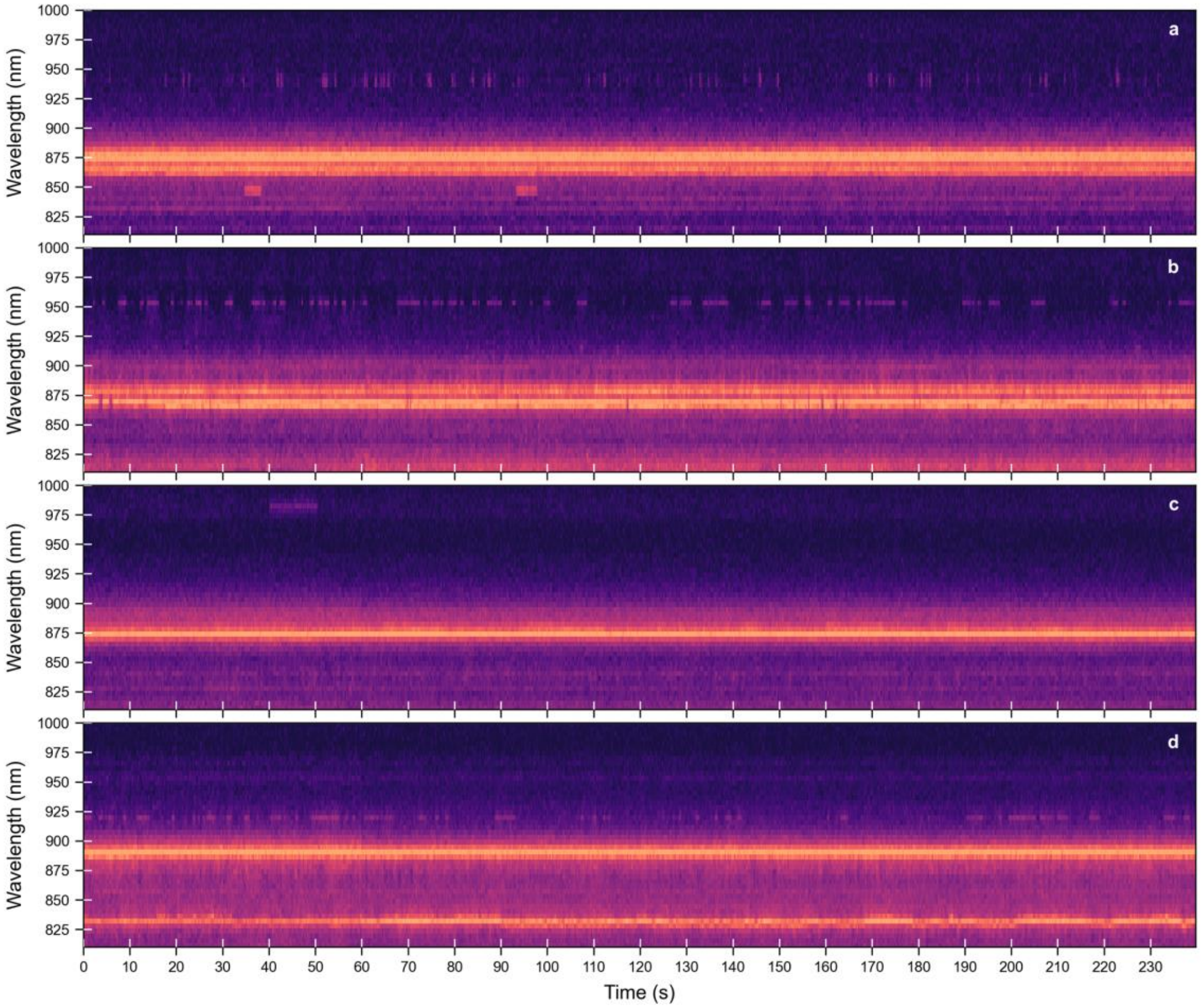

**Figure 6.** Spectrally resolved time series from four laser-written sites under 785 nm excitation. (a)–(d) Wavelength–time maps assembled from consecutive spectra with a 0.3 s frame time and 240 s of cumulative acquisition (a 60 s sequence followed by a 180 s sequence recorded

approximately one hour later). Each panel is shown on an individually normalized logarithmic intensity scale.

## 4. Discussion

### 4.1. Complementary roles of laser writing and annealing atmosphere

The comparison with our earlier argon-annealed study gives a process-level indication that the annealing environment determines the spectral distribution of the emitters formed at laser-modified sites [21]. Both studies use the same single-shot femtosecond writing platform, the same sample-preparation route and the same furnace for high-temperature annealing. After argon annealing, the emission was dominated by ZPLs between approximately 570 and 800 nm, whereas after oxygen annealing peaks above 800 nm are prevalent and extend towards 1 μm. The broader literature likewise identifies oxygen-rich processing as an important route to long-wavelength hBN emitters [16–20].

A simple physical picture is that the laser pulse and the anneal play complementary roles. Femtosecond irradiation locally generates vacancies, disorder and reactive sites, and thereby defines regions susceptible to reconstruction. High-temperature annealing enables diffusion and defect-complex formation, while the oxygen environment influences incorporation, termination, charge compensation and local bonding. The emitter population is thus controlled jointly by the initial laser modification, which sets the position, and the subsequent chemical environment, which sets the spectral range.

### 4.2. Candidate microscopic origins

The spectral diversity at individual written sites indicates that several microscopic configurations contribute. Which centres are formed or stabilized at a given site may depend not only on the irradiation conditions but also on the local material environment, including impurities, pre-existing lattice defects and strain. Local impurity concentrations can vary substantially within hBN samples [23–25], and strain can modify defect formation energies and so alter the likelihood of forming different defects [26]. In insulating hBN, different charge states provide a further source of diversity, with optical excitation and charge exchange with nearby defect states influencing the charge state of individual emitters [27]. The assignments below are tentative and are intended to indicate which known candidates are compatible with the different parts of the measured spectral range.

The pronounced emission near 575 nm under 532 nm excitation may be associated with an intradefect transition of negatively charged $V_BO_N$. This oxygen-containing defect has a small calculated Huang–Rhys factor of 0.66, a ZPL near 2 eV and a calculated radiative lifetime of 11.8 ns [28]. It is predicted to be stable with a comparatively favourable formation energy [29]. Boron-vacancy mobility at elevated temperature, together with a calculated binding energy of up to 3.6 eV for the neutral $V_BO_N$ complex [30], makes the formation of such complexes plausible under the present writing and annealing conditions.

For the NIR population, $O_NV_N$ and $O_NV_N$–H complexes are particularly relevant. They have recently been proposed as NIR single-photon emitters, with calculated ZPLs of 1.61 eV for positively charged $O_NV_N$ and 1.48 eV for neutral $O_NV_N$–H [20]. Several geometries of $O_NV_N$–H are predicted within the same spectral region, and kinetically trapped configurations could contribute to the observed spectral spread. The optically active charge states of these complexes are metastable [20], which offers one possible origin of the intermittent emission in figure 6.

Carbon-related centres may contribute to parts of the broader spectral response and have been associated with several UV-to-visible hBN emitters [10,11,23,24]. Neutral carbon trimers are calculated to have ZPLs at 1.62 eV for $C_2C_N$ and 1.69 eV for $C_2C_B$, with weak electron–phonon coupling and a characteristic phonon-sideband feature approximately 150–170 meV below the ZPL [31,32]. Similar spectral structure occurs in selected spectra, particularly at the site in figure S2(d). Carbon monomers $C_B$ may also contribute, since electron capture by positively charged $C_B$ has been predicted to produce emission near 1.70 eV [33]. Although these centres do not contain oxygen, laser-induced vacancy formation followed by high-temperature annealing could facilitate vacancy-mediated diffusion and rearrangement of native carbon impurities into larger complexes [34].

### 4.3. Scope of the present demonstration and route to single-emitter arrays

The central result of this work is positional: NIR emission in hBN, previously found only at uncontrolled locations, is activated at laser-defined coordinates with high probability. Several aspects define the scope of this demonstration. The placement precision is currently limited by the approximately 1.6 μm writing spot and can be improved by tighter focusing or by operating near the modification threshold, where subwavelength modification has been reported [8]. Many sites host more than one emitter, and photon correlations were measured on selected sites, so that the fraction of sites with a single emitter remains to be determined. The pulse-energy dependence and the multiplicity statistics both point to lower writing energies as the means of reducing site occupancy, and oxygen partial pressure, annealing temperature and duration provide further control over the density of stabilized centres. The demonstration that the spectral range can be selected by the annealing atmosphere, independently of the writing step that defines position, is what makes this optimization tractable.

## 5. Conclusion

We have shown that single-shot femtosecond laser writing followed by annealing in oxygen activates NIR emitters in exfoliated hBN at predefined coordinates. The written matrices are recovered in post-anneal PL maps, including in narrow bands centred at 800 and 900 nm, and 83% of the spectra acquired at written sites under 785 nm excitation and 71% under 635 nm excitation contain resolved emission above 810 nm, with lines extending to 1 μm. The emission intensity increases with writing-pulse energy over the studied range. Selected sites show nanosecond PL lifetimes and photon antibunching, with a raw, uncorrected $g^{(2)}(0)$ of 0.31 in the 850 nm band demonstrating single-photon emission from a laser-defined NIR site. Spectral time series show bands that persist over the recorded intervals as well as intermittent emission beyond 900 nm. Together with the visible emission obtained after argon annealing of identically written samples, these results indicate that laser writing defines where emitters form while the annealing atmosphere selects their spectral range. Reducing site occupancy through lower writing energies and optimized annealing is the remaining step towards deterministic arrays of NIR single-photon emitters in hBN.

## Data availability statement

The data that support the findings of this study are available from the corresponding author upon reasonable request.

## Acknowledgements

This work was funded by the Research Council of Lithuania, Project No. P-ITP-24-22.

## Conflict of interest

The authors declare no competing interests.

## Supplementary material

Peak-detection criteria, additional spectral time series, additional matched-site multiwavelength spectra, time-resolved PL and photon-correlation fit models and parameters, and an optical image of a laser-written matrix.